%% file: main.tex
\documentclass[10pt, conference, letterpaper]{IEEEtran}
\IEEEoverridecommandlockouts
\usepackage{amsfonts}
\usepackage{amssymb}
\usepackage{array, makecell}
\usepackage{mathtools}
\usepackage{graphicx}
\usepackage{subfigure}
\usepackage{enumerate}
\usepackage{amsmath}
\usepackage{color}
\usepackage{amsthm}
\usepackage{algorithm}
\usepackage{algpseudocode}
\usepackage{stfloats}
\usepackage{multirow}
\theoremstyle{definition}
\usepackage{color}
\allowdisplaybreaks[4]

\usepackage[bottom=1.16in,left=0.6in,right=0.6in,top=0.69in]{geometry}

\input{macros.tex}

\def\BibTeX{{\rm B\kern-.05em{\sc i\kern-.025em b}\kern-.08em
		T\kern-.1667em\lower.7ex\hbox{E}\kern-.125emX}}
		
\begin{document}	
		
\title{Federated Learning via \\
Indirect Server-Client Communications}

%

\author{\IEEEauthorblockN{Jieming Bian}
\IEEEauthorblockA{Department of ECE \\
University of Miami\\
Coral Gables, FL 33146 \\
jxb1974@miami.edu
}
\and
\IEEEauthorblockN{Cong Shen}
\IEEEauthorblockA{Department of ECE \\
University of Virginia\\
Charlottesville, VA 22904 \\
cong@virginia.edu}
\and
\IEEEauthorblockN{Jie Xu}
\IEEEauthorblockA{Department of ECE\\
University of Miami\\
Coral Gables, FL 33146 \\
jiexu@miami.edu
}}

\maketitle

\begin{abstract}
Federated Learning (FL) is a communication-efficient and privacy-preserving distributed machine learning framework that has gained a significant amount of research attention recently. Despite the different forms of FL algorithms (e.g., synchronous FL, asynchronous FL) and the underlying optimization methods, nearly all existing works implicitly assumed the existence of a communication infrastructure that facilitates the direct communication between the server and the clients for the model data exchange. This assumption, however, does not hold in many real-world applications that can benefit from distributed learning but lack a proper communication infrastructure (e.g., smart sensing in remote areas). In this paper, we propose a novel FL framework, named FedEx (short for FL via Model Express Delivery), that utilizes mobile transporters (e.g., Unmanned Aerial Vehicles) to establish indirect communication channels between the server and the clients. Two algorithms, called FedEx-Sync and FedEx-Async, are developed depending on whether the transporters adopt a synchronized or an asynchronized schedule. Even though the indirect communications introduce heterogeneous delays to clients for both the global model dissemination and the local model collection, we prove the convergence of both versions of FedEx. The convergence analysis subsequently sheds lights on how to assign clients to different transporters and design the routes among the clients. The performance of FedEx is evaluated through experiments in a simulated network on two public datasets.

\end{abstract}

\section{Introduction}
In recent years, Federated Learning (FL) has emerged as a popular distributed machine learning framework where a number of distributed clients can collaboratively train a common machine learning model under the coordination of a parameter server without exposing their own data to another party. With an unprecedented amount of data being generated on edge devices such as smart phones and Internet-of-Things (IoT) devices as well as the rising privacy concerns associated with uploading this data to the cloud, FL is now widely considered as the next-generation machine learning paradigm to power a broad variety of applications, ranging from healthcare to agriculture, transportation, industrial IoT and mobile applications due to its distributed nature and privacy-preserving advantage. 

A main aspect that makes FL stand out from other distributed learning frameworks is its specific consideration on the communication efficiency between the clients and the server. Particularly, the local stochastic gradient descent (SGD) algorithm \cite{stich2018local} and the FedAvg algorithm \cite{mcmahan2017communication} let clients perform multiple local SGD iterations on their own datasets before uploading the results to the parameter server for aggregation. Compared with earlier distributed learning algorithms such as distributed SGD \cite{balcan2012distributed} where local computation results must be uploaded to the server after every iteration, local SGD (or FedAvg) has a clear advantage in terms of the communication efficiency while still enjoying guaranteed convergence. This idea inspired many follow-up FL algorithms developed based on different optimization methods and makes FL the favorite choice in communication-constrained learning settings where either the bandwidth between the clients and the server is limited or the communication pattern is random and sporadic. 

\begin{figure}[tb]
	\centering
	\includegraphics[width=0.8\linewidth]{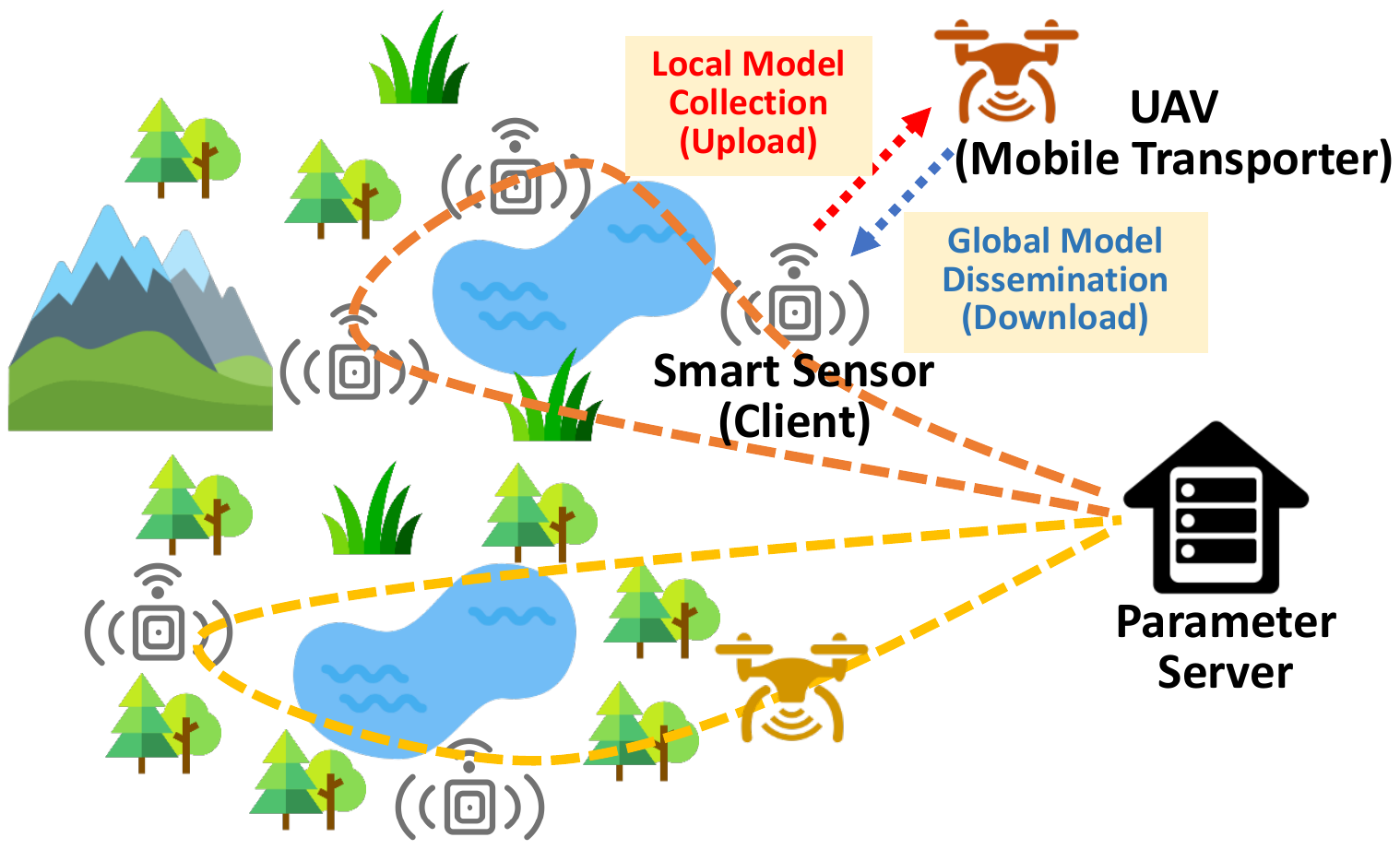}
	\caption{Illustration of FedEx applied to smart sensing in remote areas with no communication infrastructure.} \label{sensing-app}
	\vspace{-15 pt}
\end{figure}%

Despite the differences in the adopted optimization methods and the focused settings, nearly all existing works implicitly assumed that the clients can \textit{directly} communicate with the server. In the asynchronous FL category, the communication assumption is relaxed in some works \cite{avdiukhin2021federated, basu2019qsparse} so that clients may have random and sporadic communication patterns with the server, but clients can still directly communicate with the server, albeit less regularly. However, in many real-world systems, clients may not be able to directly communicate with the server at all due to the lack of a proper communication infrastructure. 
It is thus imperative to understand whether FL can still work without direct server-client communications and how to optimize the FL algorithms in these settings. 

In this paper, we propose a new FL framework, called FedEx (Federated Learning via Model Express Delivery), for the considered FL system with no direct server-client communications. To address the no direct communication challenge, FedEx devises an active mobility mechanism, which utilizes mobile transporters (e.g., Unmanned Aerial Vehicles / UAVs) to establish \textit{indirect} communication channels between the server and the clients to facilitate the model information exchange. In other words, the mobile transporters serve as an intermediary between the server and the clients to disseminate global models and collect local model updates, analogous to delivery trucks in a traditional parcel express delivery system. Fig. \ref{sensing-app} illustrates the application of FedEx to smart sensing in remote areas with no communication infrastructure. 

However, the indirect communication also brings significant new challenges to the convergence analysis and optimization of FedEx. First, in both the global model dissemination phase and the local model collection phase, delay is inevitably introduced by the indirect communication as it takes time for the mobile transporters to move from one location to another. It is unclear whether FedEx can still converge under this delay and if so, how fast. Second, depending on the transporter scheduling policy, learning can be either synchronized or asynchronized at the transporter level, thereby further complicating the convergence analysis of FedEx. Third, clients experience heterogeneous delays depending on their locations and the routes chosen by the transporters. Thus, the performance of FedEx is also contingent on how clients are assigned to the transporters and how the transporters design their routes. We summarize the main contributions of this paper below.
\begin{itemize}
    \item To our best knowledge, we propose the first FL framework via indirect server-client communications. Two algorithms, coined FedEx-Sync and FedEx-Async, are proposed depending on whether the transporters synchronize their tours among the assigned clients. 
    \item We prove the convergence of both FedEx-Sync and FedEx-Async using a virtual sequence technique. 
    \item Based on the specific forms of the convergence bounds, a bi-level optimization algorithm is proposed to solve the joint client assignment and route design problem. 
    \item The experiments results using two public datasets validate the efficacy of FedEx and are consistent with our theory. 
\end{itemize}

\section{Problem Formulation}
We consider an FL system with one parameter server and $N$ clients, which are distributed over a large area \textit{without} direct communication capabilities. In other words, no client has a direct communication channel with the server and any pair of clients do not communicate with each other. For notation simplicity, we index the server as $0$ and the clients by the set $\mathcal{N} = \{1, 2, \cdots, N\}$. The server and the clients are deployed in fixed locations and do not move. Given the location coordinates of the server and the clients, one can easily calculate the (symmetric) distance matrix $D \in \mathbb{R}^{(N+1)\times (N+1)}$ that describes the distance between any two devices. Specifically, $D_{0i}=D_{i0}$ is the distance between the server and client $i$ and $D_{ij} = D_{ji}$ is the distance between clients $i$ and $j$. 

Each client $i$ has a dataset and the clients must together train a machine learning model under the coordination of the server by solving the following distributed optimization problem:
\begin{align}
    \min_x f(x) = \frac{1}{N}\sum_{i=1}^N f_i(x) = \frac{1}{N}\sum_{i=1}^N \mathbb{E}_{\zeta_i}[F_i(x, \zeta_i)],
\end{align}
where $f_i: \mathbb{R}^d \to \mathbb{R}$ is a non-convex loss function for client $i$, $F_i$ is the estimated loss function based on a mini-batch data sample $\zeta_i$ drawn from client $i$'s dataset and $x \in \mathbb{R}^d$ is the model parameter to learn. 

To train such a machine learning model, conventional FL frameworks require periodic/non-periodic communications between the clients and the server. However, in our considered setting, all direct communication channels between the clients and the server are absent and hence, these FL frameworks all fail to work. In the next section, we propose a novel FL framework to enable FL in such extreme communication scenarios (i.e., no direct communications). 

\section{FedEx: FL via Model Express Delivery}
To address the lack of direct communications between the clients and the server, our idea is to leverage mobile transporters (e.g., UAVs) to build indirect communication channels between the server and the clients. These mobile transporters transport global/local models between the server and the clients, just like delivery trucks transport parcels between the warehouse and the customers. We call this new FL framework FedEx, short for FL via Model Express Delivery. 

Consider that $K$ mobile transporters can be used in FedEx. The specific value of $K$ depends on the available system resources and we consider it as given in this paper. Without loss of generality, we assume that all clients have the same computing speed and all transporters have the same moving speed. We discretize time into slots, indexed by $t=0, 1, 2, \cdots$, where each time slot corresponds to the duration of completing one local training step by a client. Furthermore, with a slight abuse of notation, we let the matrix $D$ represent the time (in terms of time slots) needed for the mobile transporter to travel from one location to another instead of the distance to simplify our notations. 

FedEx works by first partitioning the clients into $K$ non-overlapping subsets and assigning each subset to one mobile transporter. Let $\mathcal{R}_k \subset \mathcal{N}$ represent the subset of clients that are covered by transporter $k$. We have $\mathcal{R}_k\cap\mathcal{R}_{k'} = \emptyset, \forall k \neq k'$ and $\cup_{k=1}^K \mathcal{R}_k = \mathcal{N}$. Moreover, let $R_k = |\mathcal{R}_k|$ be the number of clients in subset $\mathcal{R}_k$. For each transporter $k$, it determines a round-trip tour among the server and the clients in $\mathcal{R}_k$. Let $z_{ij}, \forall i,j\in \mathcal{R}_k\cup\{0\}$ be a binary variable indicating whether a path 
from device $i$ to device $j$ is included in the tour, then the round-trip time (RTT) can be easily calculated as
\begin{align}
    \Delta_k = \sum_{i=0}^N \sum_{j\neq i, j=0}^N D_{ij}z_{ij}.
\end{align}
To make sure that the tour covers all devices exactly once, we have the constraints that each device has exactly one incoming path and one outgoing path, which can be expressed as $2(N+1)$ linear equations:
\begin{align}
    \sum_{i=0, i\neq j}^N z_{ij} = 1, \sum_{j=0, j\neq i}^n z_{ij} = 1, \forall j = 0, \cdots, N.
\end{align}
Finding the shortest tour for a given set of clients $\mathcal{R}_k$ is essentially a {\it travelling salesman problem} and we defer the optimization of client assignment among the transporters to Section \ref{opt_section} after understanding the convergence behavior of FedEx. For now, we treat $\mathcal{R}_k, \forall k = 1, \cdots, K$ as already decided along with the corresponding tour RTT $\Delta_k$. 



\subsection{FedEx-Sync}
In the synchronized version of FedEx, namely FedEx-Sync, the transporters depart from the server \textit{at the same time} every time they start a new tour among their assigned clients. Because the transporters have different tour RTTs, the ones with shorter RTT need to wait for the others with longer RTT to come back to start the next tour. Therefore, FedEx-Sync is naturally composed of synchornized learning rounds, with each round having $\Delta \triangleq \max_{k}\Delta_k$ time slots. In each round (denote the first slot of this round as $t_0$), the following events occur.
\begin{itemize}
    \item At the beginning of each round, each transporter downloads the current global model $x^{t_0}$ from the server. The transporters then start a tour among their assigned clients according to the pre-determined client visiting order.
    \item  When a transporter (say transporter $k$) meets a client (say client $i$) at time $t > t_0$, client $i$ downloads the global model, i.e., $x^{t_0}$, that transporter $k$ currently carries. Then the transporter leaves and client $i$ uses $x^t_k = x^{t_0}$ as the initial model to train a new local model using its own local dataset until the next time it meets the transporter. Because the transporter takes $\Delta$ time slots to revisit client $i$, the local training will last $\Delta$ time slots. The local training uses a mini-batch SGD method:
    \begin{align}
        x^{s+1}_i = x^s_i - \eta g^s_i, \forall s = t, \cdots, t + \Delta-1,
    \end{align}
    where $g^s_i = \nabla F_i(x^s_i, \zeta^s_i)$ is the stochastic gradient on a randomly drawn mini-batch $\zeta^s_i$ and $\eta$ is the learning rate. Let $m^t_i \in \mathbb{R}^d$ be the cumulative local updates (CLU) of client $i$ at time $s$ since its last meeting with the transporter, which is updated recursively as follows
    \begin{align}
        m^{s}_i = \sum_{s'=t}^{s-1} \eta g^s_i, \forall s = t, \cdots, t+\Delta.
    \end{align}
    \item When a transporter (say transporter $k$) meets a client (say client $i$) at time $t > t_0$, client $i$ also uploads its current CLU to transporter $k$. Note, however, that this CLU is obtained based on the global model from the \textit{previous} round, i.e., $x^{t_0 - \Delta}$. Transporter $k$ maintains an aggregated CLU $u^t_k$ during the current tour to save storage space and updates it whenever a new client CLU is received according to
    \begin{align}
        u^t_k = u^{t-1}_k + m^t_i.
    \end{align}
    \item When the transporter returns to the server, the aggregated CLU is used to update the global model. In FedEx-Sync, the global model is updated synchronously at the end of each round as follows
    \begin{align}
        x^{t_0+\Delta} = x^{t_0+\Delta - 1} - \frac{1}{N}\sum_{k=1}^K u^{t_0+\Delta - 1}_k.
    \end{align}
\end{itemize}


\subsection{FedEx-Async}
The synchronization in FedEx-Sync is achieved by asking faster transporters to wait for slower transporters. This, however, introduces extra delays for faster transporters. In the case where the slowest transporter takes a tour with a very large RTT, then all the other transporters will have to wait for a long time before starting their next tour. In FedEx-Async, we remove such waiting time by letting the transporter start a new tour immediately after finishing the previous tour. In this way, more clients will be able to perform more frequent global/local model exchanges with the server. FedEx-Async share many similarities with FedEx-Sync and the biggest difference is that each transporter will have \textit{individualized} learning rounds not necessarily synchronized with others. For transporter $k$, its learning round lasts $\Delta_k$ time slots and the following events occur in each round (denote the first slot as $t_0$).
\begin{itemize}
    \item At the beginning of each round, transporter $k$ downloads the current model $x^{t_0}$ from the server. Then it starts its tour among the assigned clients.
    \item When transporter $k$ meets client $i$ at time $t>t_0$, client $i$ downloads $x^{t_0}$ from transporter $k$ and uses $x^t_k = x^{t_0}$ as the initial model to train a new local model until the next time it meets the transporter. Different from FedEx-Syncs, the local training will last $\Delta_k$ time slots, which are different across transporters. 
    \item When transporter $k$ meets client $i$ at time $t > t_0$, client $i$ also uploads its current CLU, which is obtained based on the previous round global model $x^{t_0 - \Delta_k}$, to transporter $k$. Transporter $k$ then updates its aggregated CLU $u^t_k$. 
    \item When the transporter returns to the server, the global model is updated as follows
    \begin{align}
        x^{t_0+\Delta_k} = x^{t_0+\Delta_k - 1} - \frac{1}{N} u^{t_0+\Delta_k - 1}_{k}.
    \end{align}
    Again, this is different from FedEx-Sync since the server does not have to wait for all transporters to update the global model. Note that it is possible that multiple transporters can return to the server in the same time slot (say $t$). In this case, the global model update rule is changed to
    \begin{align}
        x^{t_0+\Delta_k} = x^{t_0+\Delta_k - 1} - \frac{1}{N}\sum_{k' \in \mathcal{S}^{t_0+\Delta_k}} u^{t_0+\Delta_k - 1}_{k'} ,
    \end{align}
    where $\mathcal{S}^{t_0+\Delta_k -1}$ is the set of clients that return to the server at time slot $t_0+\Delta_k -1$. 
\end{itemize}

\section{Convergence Analysis}
In this section, we analyze the convergence of FedEx. Because FedEx-Sync can be considered as a special case of FedEx-Async where all $\Delta_k, \forall k$ take the same value, we will focus on the convergence analysis of FedEx-Async. 

\subsection{Aligning Client Training}
Before analyzing the convergence of FedEx, we first describe an equivalent view of FedEx that aligns the local training of clients in the same subset. Consider a learning round of transporter $k$ that carries the global model $x^{t_0}$. Because of the different locations of clients in $\mathcal{R}_k$, the clients receive $x^{t_0}$ and start their new round of local training at different time slots. Once they finish the current round of local training, their CLUs based on $x^{t_0}$ will be uploaded via the transporter to the server at time slot $t_0 + 2\Delta_k$. At that moment, the global model gets an update using these CLUs. 

The unaligned local training of clients, even if covered by the same transporter, would create a major challenge for the convergence analysis of FedEx. Fortunately, there is an equivalent (but imaginary) client training procedure that produces exactly the same global model sequence. Specifically, imagine that clients in $\mathcal{R}_k$ receive the global model $x^{t_0}$ immediately at time slot $t_0$ and perform their local training for  $\Delta_k$ time slots. Then their CLUs are delayed one round to be uploaded to the server. That is, at time slot $t_0 + 2\Delta_k$, the global model gets an update. 
It is clear that the global model update is not affected at all by this change but the local training among clients in the same subset $\mathcal{R}_k$ is now perfectly aligned. Since we are interested in the convergence of the global model, we consider the equivalent aligned client training procedure in our convergence proof. Where the local training of the clients is aligned while the global model evolution is unaffected. Essentially, the alignment moves the download delay to the upload phase, but the total delay remains the same. With this change, FedEx-Sync becomes a familiar synchronous FL algorithm but with one round CLU upload delay. In addition to the CLU upload delay, Fed-Async still features asynchronized learning across the client subsets.


\subsection{Assumptions}
Our convergence analysis will utilize the following standard assumptions. 
\begin{assumption}[Lipschitz Smoothness]
There exists a constant $L > 0$ such that $\|\nabla f_i(x) - \nabla f_i(y)\|\leq L \|x - y\|$, $\forall x, y \in \mathbb{R}^d$ and $\forall i = 1, \cdots, N$. 
\end{assumption}

\begin{assumption}[Unbiased Local Gradient Estimate]
The local gradient estimate is unbiased, i.e., $\mathbb{E}_\zeta F_i(x, \zeta) = \nabla f_i(x)$, $\forall x$ and $\forall i = 1, \cdots, N$. 
\end{assumption}

\begin{assumption}[Bounded Gradient]
There exists a constant $G > 0$ such that $\mathbb{E}\|\nabla F_i(x, \zeta)\|^2 \leq G^2$, $\forall x \in \mathbb{R}^d$ and $\forall i = 1, \cdots, N$. 
\end{assumption}

\begin{assumption}[Bounded Variance]
There exists a constant $\sigma > 0$ such that $\mathbb{E}_{\zeta}\|\nabla F_i(x, \zeta) - \nabla f_i(x)\|^2\leq \sigma^2$, $\forall x \in \mathbb{R}^d$ and $\forall i = 1, \cdots, N$. 
\end{assumption}
 
\subsection{Convergence Bound}
Our convergence analysis relies on understanding the relationship between and the evolution of two sequences of the global model. The \textit{real sequence} of the global model is the actual global models maintained at the server over time, which can be calculated as follows according to FedEx:
\begin{align}
    x^t = x^0 - \frac{1}{N}\sum_{i=1}^N \sum_{s=0}^{\phi_i(t)}\eta g^s_i, \forall t,
\end{align}
where we define $\phi_i(t)$ as the time slot up to when all corresponding gradients of client $i$ have been received at time $t$. In other words, at time slot $t$, the server has received gradients $g^0_i, \cdots, g^{\phi_i(t)}_i$ from client $i$ (via the transporter). In FedEx-Sync, all clients have the same indirect communication patterns with the server and hence $\phi_i(t) = \phi_j(t), \forall i,j \in \mathcal{N}$. In FedEx-Async, clients belonging to the same transporter have the same indirect communication patterns with the server and hence, $\phi_i(t) = \phi_j(t), \forall i, j\in \mathcal{R}_k, \forall k$. 

The \textit{virtual sequence} of the global model is defined in the imaginary case where all client gradients are uploaded to the server immediately after they have been calculated. Similar virtual sequences have been utilized in  \cite{yuan2016convergence}. However, our convergence proof is tailored to the specific problems in our paper and different than all prior works. Specifically, the virtual sequence is defined as
\begin{align}
    v^t = x^0 - \frac{1}{N}\sum_{i=1}^N \sum_{s=0}^{t-1}\eta g^s_i, \forall t.
\end{align}

Clearly, there is a discrepancy between the real sequence and the virtual sequence due to the delayed upload of the client gradients. For FedEx-Sync, this delay is bounded by
\begin{align}
    (t-1) - \phi_i(t) \leq 2 \Delta, \forall i \in \mathcal{N}.
\end{align}
For FedEx-Async, the delay is bounded by
\begin{align}
    (t-1) - \phi_i(t) \leq 2 \Delta_k, \forall i \in \mathcal{R}_k, \forall k.
\end{align}
\begin{lemma}
The difference between the real global model and the virtual global model can be bounded as follows
\begin{align}
    \mathbb{E}\|v^t - x^t\|^2 \leq \frac{4\eta^2 G^2}{N}\sum_{k=1}^K R_{k} \Delta^2_{k}.
     \label{r-v-global-bound}
\end{align} 
The average difference between all clients' local models and the virtual global model is bounded as follows
\begin{align}
    \frac{1}{N}\sum_{i=1}^N \mathbb{E}\|v^t - x^t_i\|^2 \leq \frac{18\eta^2 G^2}{N}
    \sum_{k=1}^K R_{k} \Delta^2_{k}.
\end{align}
\end{lemma}

\begin{theorem}
By setting the learning rate $0 < \eta \leq 1/L$, we have
\begin{align}
    &\frac{1}{T}\sum_{t=0}^{T-1}\mathbb{E}\|\nabla f(x^t)\|^2  \nonumber\\
    \leq& \frac{4}{\eta T} (f(x^0) - f^*) 
    + \frac{44 \eta^2 G^2 L^2}{N}\sum_{k=1}^K R_k \Delta^2_k + \frac{2L\eta\sigma^2}{N}.
\end{align}
\end{theorem}
\begin{remark}
For $T \geq N^3$, by setting the learning rate as $\eta = \frac{\sqrt{N}}{L\sqrt{T}}$, the convergence bound recovers the same $O(\frac{1}{\sqrt{NT}})$ convergence rate of the classical synchronous FL \cite{yu2019parallel}. \end{remark}

\begin{remark}
For FedEx-Sync, the convergence bound can be tightened a little bit because the real sequence and the virtual sequence periodically coincide with each other. In addition, because the effective RTT of all transporters is $\max_k \Delta_k$, the convergence bound reduces to
\begin{align}
    \frac{2}{\eta T} (f(x^0) - f^*) 
    + 18 \eta^2 G^2 L^2 (\max_k\Delta_k)^2 + \frac{L\eta\sigma^2}{N}.
\end{align}
\end{remark}


\section{Client Assignment and Route Design} \label{opt_section}
In this section, we study the joint client assignment and route design problem to optimize the convergence bound of FedEx. 

\subsection{Problem Formulation}
We consider a typical setting where the number of clients is much larger than the number of transporters, i.e., $N \gg K$, due to the limited transporter availability and potentially massive deployment of IoT devices. Let $a_i \in \{1, \cdots, K\}$ be the assignment variable of client $i$, indicating which transporter it is assigned to. We also collect the assignment variables of all clients in $\a = (a_1, \cdots, a_N)$. Clearly, $\mathcal{R}_k = \{i: a_i = k\}$. 

Given the assigned clients $\mathcal{R}_k$ for each transporter $k$, we can design a route to minimize the RTT. Let $\Delta_k(\mathcal{R}_k)$ be the minimum RTT given a set of client $\mathcal{R}_k$. Alternatively, $\Delta_k(\mathcal{R}_k)$ can also be written as $\Delta_k(\a)$ since $\mathcal{R}_k$ is determined by $\a$. Client assignment is to solve the following optimization problems
\begin{align}
    \text{FedEx-Sync}: &\min_\a \max_k\Delta_k(\a),\\
    \text{FedEx-Async}: &\min_\a \sum_{k} R_k(\a) \Delta^2_k(\a).
\end{align}
The above problem is a difficult combinatorial optimization problem. Next, we propose a new algorithm to solve this problem. 

\subsection{Bi-level Optimization}
We develop a bi-level optimization algorithm, called CARD (short for Client Assignment and Route Design). 

\subsubsection{Inner-level optimization} The inner-level optimization is to solve the minimum RTT given a set of client $\mathcal{R}_k$, for each transporter $k$, namely computing $\Delta_k(\a)$ for a given $\a$. This is a classical traveling salesman problem (TSP) \cite{lin1973effective, flood1956traveling, junger1995traveling}. Considering the high complexity of dynamic programming (i.e., $O(2^n n^2)$ where $n$ is the number of nodes), we use a heuristic algorithm 2-OPT \cite{croes1958method}, which has a time complexity of $O(n^2)$, to compute $\Delta_k(\a)$ for a given $\a$ in our implementation.


\subsubsection{Outer-level optimization} To solve the outer-level optimization problem to determine the optimal client assignment, we resort to Gibbs Sampling. For notation simplicity, we use a unified cost function $C(\a)$, which equals $\max_k \Delta_k(\a)$ for FedEx-Sync and $\sum_{k} R_k(\a) \Delta^2_k(\a)$ for FedEx-Async. The CARD algorithm visits each client according to a pre-defined sequence, generates a probability distribution of its assignment decision while holding other clients' assignment decision unchanged, and samples a new assignment decision according to this distribution. Repeating this process for sufficiently many iterations ensures that the assignment converges to the optimal solution with high probability. 



\section{Experiments}
In this section, we  evaluate the performance of FedEx using a simulated network environment and standard public datasets. 

\subsection{Experiment Setup}
\begin{figure*}[tt]
    \centering
    \begin{minipage}[t]{0.32\linewidth}
        \includegraphics[width=1\linewidth]{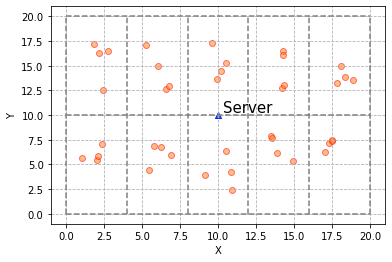}
        \vspace{-20 pt}
        \caption{Network Illustration}
	    \label{fig:clients}
    \end{minipage}
    \begin{minipage}[t]{0.32\linewidth}
        \includegraphics[width=1\linewidth]{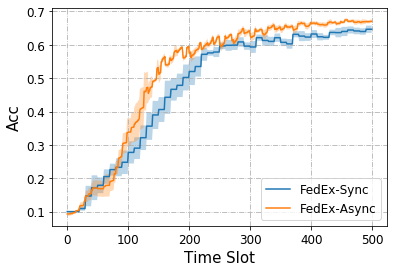}
        \vspace{-20 pt}
        \caption{FedEx under i.i.d data on FMNIST}
	    \label{fig:fmnist_iid}
    \end{minipage}
    \begin{minipage}[t]{0.32\linewidth}
        \includegraphics[width=1\linewidth]{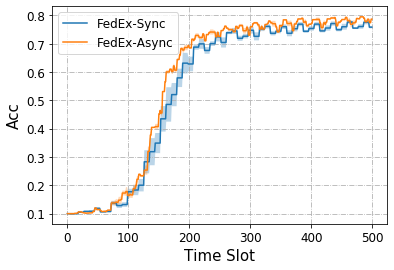}
        \vspace{-20 pt}
        \caption{FedEx under i.i.d data on SVHN}
	    \label{fig:svhn_iid}
    \end{minipage}    
\end{figure*}

\textbf{Network}. We simulate a network with no direct communications where one parameter server and 40 clients are distributed over an area as shown in Fig. \ref{fig:clients}. The whole area is first divided into 10 blocks of equal size (i.e., 4 units $\times$ 10 units), and 4 clients are randomly distributed in each block. We simulate $K = 4$ mobile transporters in most of our experiments, which have the same moving speed of 4 units per time slot.  

\textbf{DNN Model and Dataset}. We conduct the FL experiments on two public datasets, i.e., FMNIST \cite{xiao2017fashion} and SVHN \cite{netzer2011reading}. For both datasets, we utilize LeNet \cite{lecun1998gradient} as the backbone model. 

\textbf{Dataset Splitting}. For the FMNIST experiments, each client possesses 60 training data samples. For the SVHN experiments, each client possesses 800 training data samples. Two data distributions are simulated. 
\begin{itemize}
    \item IID: The clients' data distributions are i.i.d.
    \item Non-IID: We use the Dirichlet method to create non-i.i.d. datasets, which is widely applied in FL research, e.g., \cite{chen2020fedbe}. We use a Dirichelet distribution with parameter 0.3 in the FMNIST experiments and 0.5 in the SVHN experiments. 
\end{itemize}

\textbf{Transporter Routes}. We evaluate different transporter route designs that uses different objective functions to solve the client assignment problem:
\begin{itemize}
    \item Min-Max: $\max_k \Delta_k(\a)$. 
    \item Sum-of-Weighted-Squared (SWS):  $\sum_k R_k(\a)\Delta^2_k(\a)$.
    \item Shortest-Total: $\sum_k \Delta_k(\a)$.
\end{itemize}

\subsection{Results}
\textbf{FedEx-Sync v.s. FedEx-Async under i.i.d. data}. We first compare the performance of FedEx-Sync (with Min-Max routes) and FedEx-Async (with SWS routes) under i.i.d. data. Fig. \ref{fig:fmnist_iid} and Fig. \ref{fig:svhn_iid} demonstrate their convergence curves on FMNIST and SVHN, respectively. As can be seen, under i.i.d. data, both algorithms converge but FedEx-Async outperforms FedEx-Sync. This makes sense since FedEx-Sync spends extra time in waiting for the slowest mobile transporters. 

\begin{figure*}
    \centering
    \begin{minipage}[t]{0.24\linewidth}
        \includegraphics[width=1\linewidth]{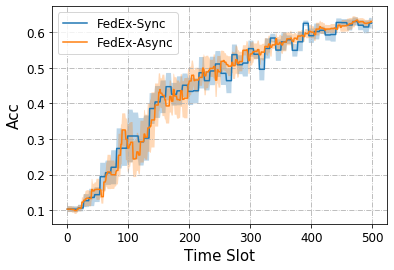}
        \vspace{-20 pt}
        \caption{non-i.i.d on FMNIST}
	    \label{fig:fmnist_type1}
    \end{minipage}   
    \begin{minipage}[t]{0.24\linewidth}
        \includegraphics[width=1\linewidth]{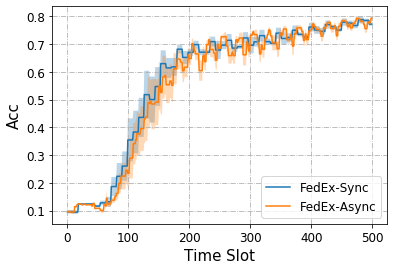}
        \vspace{-20 pt}
        \caption{non-i.i.d on SVHN}
	    \label{fig:svhn_type1}
    \end{minipage}
    \begin{minipage}[t]{0.24\linewidth}
        \includegraphics[width=1\linewidth]{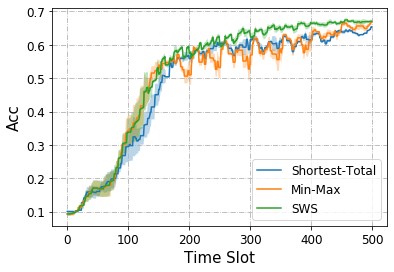}
        \vspace{-20 pt}
        \caption{Impact of Routes (Async)}
	    \label{fig:asyn-path}
    \end{minipage}
    \begin{minipage}[t]{0.24\linewidth}
        \includegraphics[width=1\linewidth]{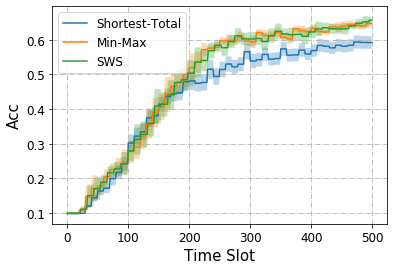}
        \vspace{-20 pt}
        \caption{Impact of Routes (Sync)}
	    \label{fig:syn-path}
    \end{minipage}    
\end{figure*}

\textbf{FedEx-Sync v.s. FedEx-Async under non-i.i.d. data}. Next, we compare FedEx-Sync and FedEx-Async non-i.i.d. data. Fig. \ref{fig:fmnist_type1} and Fig. \ref{fig:svhn_type1} show that FedEx-Sync performs similarly to FedEx-Async. Although individual clients' data can be very non-i.i.d., the overall client data of a transporter can still be similar to the entire data distribution when there are sufficiently many clients on the transporter's route.

\textbf{Impact of Routes}. We empirically investigate the impact of different transporter route designs. Fig. \ref{fig:asyn-path} compares the convergence of FedEx-Async under different route designs. The result is consistent with our theoretical analysis in Theorem 1 that the SWS design achieves the best convergence performance. Fig. \ref{fig:syn-path} shows the results for FedEx-Sync. In this case, the longest individual RTT obtained in  SWS is the same as that obtained in Min-Max, and hence SWS is also another solution of Min-Max. Therefore, Min-Max routes and SWS routes achieve similar convergence performance and outperform Shortest-Total, in accordance to the theoretical analysis in Theorem 1. 


\section{Conclusion}
In this paper, we have developed a new FL framework via indirect server-client communications to support distributed machine learning in scenarios without a communication infrastructure. Two novel algorithms have been proposed that utilize mobile transporters to disseminate global models and collect local models via device-to-device communications. We have carried out a novel convergence analysis of these algorithms under arbitrary transporter routes, for non-convex loss functions and non-i.i.d. data distributions. The result offers a principled guideline for the joint client assignment and route design. 
\bibliographystyle{IEEEtran}
\bibliography{reference}


\end{document}

%% file: macros.tex
\newcommand{\bp}{\begin{proof} \small }
\newcommand{\ep}{\end{proof} \normalsize}
\newcommand{\epx}{\end{proof} \small}
\newcommand{\bpa}{\begin{proofappx} \footnotesize }
\newcommand{\epa}{\end{proofappx} \small }
\newtheorem{theorem}{Theorem}

\newtheorem{lemma}{Lemma}
\newtheorem{assumption}{Assumption}

\newtheorem{remark}{Remark}
\newtheorem*{theorem*}{Theorem}
\newtheorem*{proposition*}{Proposition}
\newtheorem*{corollary*}{Corollary}
\newtheorem*{lemma*}{Lemma}
\newtheorem*{assumption*}{Assumption}
\newtheorem*{definition*}{Definition}
\newtheorem*{claim*}{Claim}
\newtheorem*{remark*}{Remark}

\newcommand{\be}{\begin{equation}}
\newcommand{\ee}{\end{equation}}
\newcommand{\bs}{\begin{subequations}}
\newcommand{\es}{\end{subequations}}
\newcommand{\bq}{\begin{eqnarray}}
\newcommand{\eq}{\end{eqnarray}}
\newcommand{\bqn}{\begin{eqnarray*}}
\newcommand{\eqn}{\end{eqnarray*}}

\newcommand{\ba}{\left[ \begin{array}}
\newcommand{\ea}{\\ \end{array} \right]}
\newcommand{\ben}{\begin{enumerate}}
\newcommand{\een}{\end{enumerate}}

\def\a{{\boldsymbol{a}}}

\def\real{{\mathchoice%
{\hbox{\rm\setbox1=\hbox{I}\copy1\kern-.45\wd1 R}}
{\hbox{\rm\setbox1=\hbox{I}\copy1\kern-.45\wd1 R}}
{\hbox{\scriptsize\rm\setbox1=\hbox{I}\copy1\kern-.45\wd1 R}}
{\hbox{\scriptsize\rm\setbox1=\hbox{I}\copy1\kern-.45\wd1 R}}}}

\def\Zint{{\mathchoice{\setbox1=\hbox{\sf Z}\copy1\kern-.75\wd1\box1}
{\setbox1=\hbox{\sf Z}\copy1\kern-.75\wd1\box1}
{\setbox1=\hbox{\scriptsize\sf Z}\copy1\kern-.75\wd1\box1}
{\setbox1=\hbox{\scriptsize\sf Z}\copy1\kern-.75\wd1\box1}}}
\newcommand{\complex}{ \hbox{\rm C\kern-0.45em\rule[.07em]{.02em}{.58em}%
\kern 0.43em}}

\makeatletter
\newcommand{\algmargin}{\the\ALG@thistlm}
\makeatother
\newlength{\whilewidth}
\algdef{SE}[parWHILE]{parWhile}{EndparWhile}[1]
{\parbox[t]{\dimexpr\linewidth-\algmargin}{%
		\hangindent\whilewidth\strut\algorithmicwhile\ #1\ \algorithmicdo\strut}}{\algorithmicend\ \algorithmicwhile}%
\algnewcommand{\parState}[1]{\State%
	\parbox[t]{\dimexpr\linewidth-\algmargin}{\strut #1\strut}}

\ifodd 1

\else

\fi
